\documentstyle[11pt,twoside,newpasp,epsf]{article}
\def\edcomment#1{\iffalse\marginpar{\raggedright\sl#1\/}\else\relax\fi}
\reversemarginpar

\newcommand{\mem}[1]{\mathrm{ #1}}
\newcommand{\n}{\mem{n}}
\newcommand{\msun}{\, {\rm M}_\odot}
\newcommand{\czw}{^{12}\mem{C}}
\newcommand{\cdr}{^{13}\mem{C}}

\newcommand{\nvi}{^{14}\mem{N}}
\newcommand{\fne}{^{19}\mem{F}}
\newcommand{\oac}{^{18}\mem{O}}
\newcommand{\cvi}{^{14}\mem{C}}
\newcommand{\nfu}{^{15}\mem{N}}
\newcommand{\spr}{\mbox{$s$-process}}

\begin{document}
\title{The \mbox{$s$-process} in rotating AGB stars}
 \author{Falk Herwig}
\affil{University of Victoria, Victoria, BC, Canada;
  fherwig@uvastro.phys.uvic.ca}
\author{Norbert Langer}
\affil{Universiteit Utrecht, Utrecht, The Netherlands; N.Langer@astro.uu.nl}
\author{Maria Lugaro}
\affil{IoA, University of Cambridge, Cambridge, UK; mal@ast.cam.ac.uk}

\begin{abstract}
We discuss the occurrence of the s-process during the radiative
interpulse phase of rotating AGB stars. Due to differential rotation,
protons are mixed into $\czw$-rich layers after thermal pulses,
in the course of the so called third
dredge up episode. We follow the time evolution of key isotope
abundances in the relevant layers with a post-processing code
which includes time dependant mixing and nucleosynthesis.
In rotating AGB models, the mixing persists during the entire
interpulse phase due to the steep gradient of angular velocity at the
envelope-core interface. As the layers containing protons and
$\czw$, which are formed this way, become hotter, a $\cdr$-pocket
is formed in a natural way. However,
in this situation also $\nvi$ is formed and spread over the
entire $\cdr$-pocket. We include the neutron consuming
$\nvi$(n,p) reaction in our network and determine to what extent
it reduces the production of trans-iron elements.
We propose that rotation may be responsible for the spread of
efficiencies of the $\cdr$ neutron source as required by observations.
\end{abstract}

Current models of the \spr\ in Asymptotic Giant Branch (AGB) stars
consider some kind of partial mixing of the H-rich convective envelope and
the $\czw$-rich intershell/core region, e.g.\ due to convective overshoot
(Herwig, 2000), at the end of the third dredge up phase after
advanced thermal pulses (e.g.\ Gallino etal., 1998).
Neutrons are subsequently released by the
$\cdr(\alpha,\n)$-reaction under radiative conditions
 during the interpulse phase between thermal pulses and no further mixing
 processes have been considered in this layer.

Models of rotating AGB stars display large gradients in angular
velocity at the core-envelope interface where they are able to produce a
$\cdr$- and $\nvi$-pocket next to each other. In this case
weak but persistent mixing of elements and angular momentum
continues throughout the interpulse period (Langer etal., 1999).
We have computed post-processing
nucleosynthesis models of one thermal pulse cycle from a rotating
3$\msun$ (Z=0.02) AGB star (Langer etal., 1999) \textbf{[CASE 1]}
using a new fully implicit code which treats
mixing and all relevant nuclear reactions simultaneously.
\begin{figure}
\label{fig1}
\plottwo{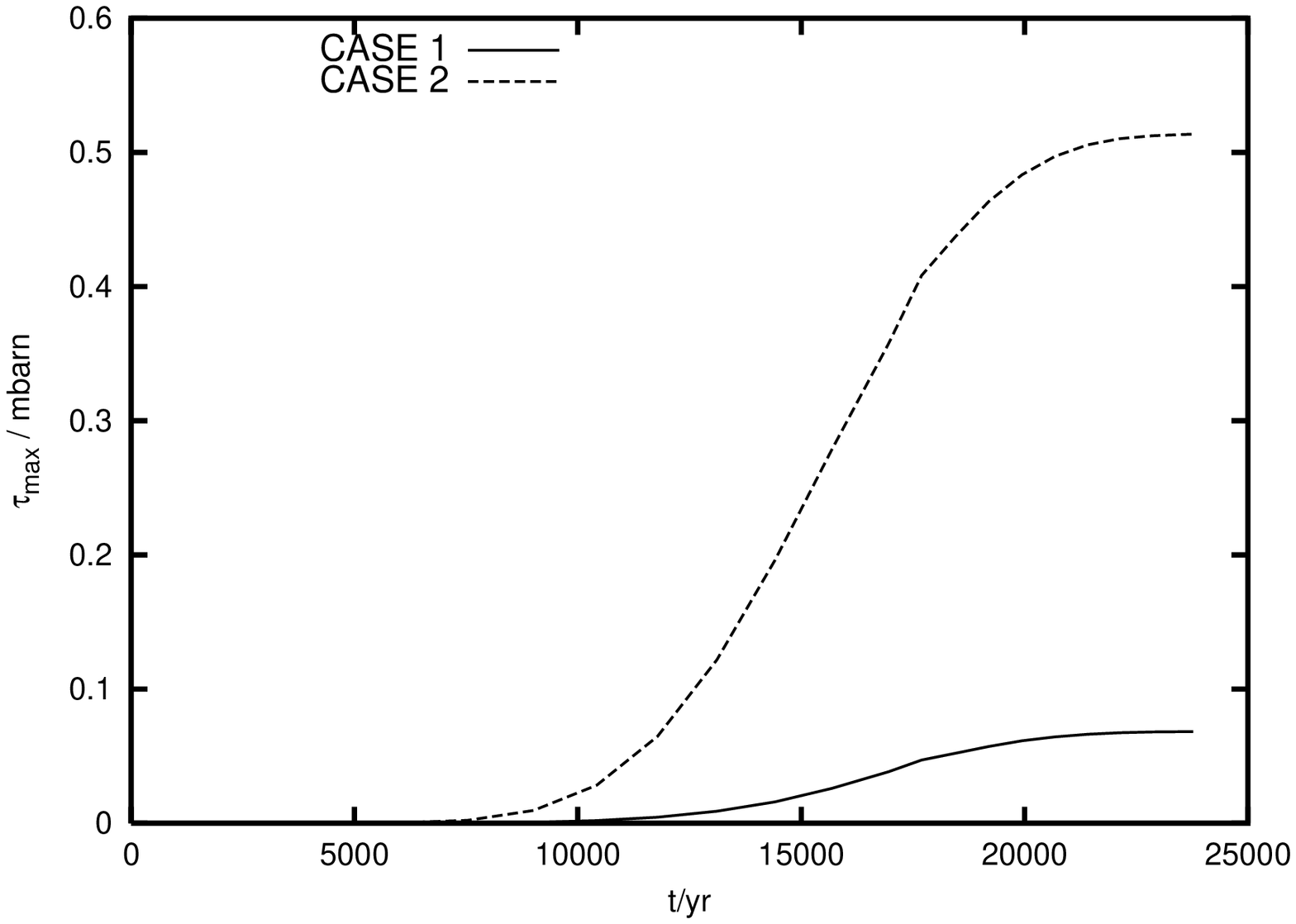}{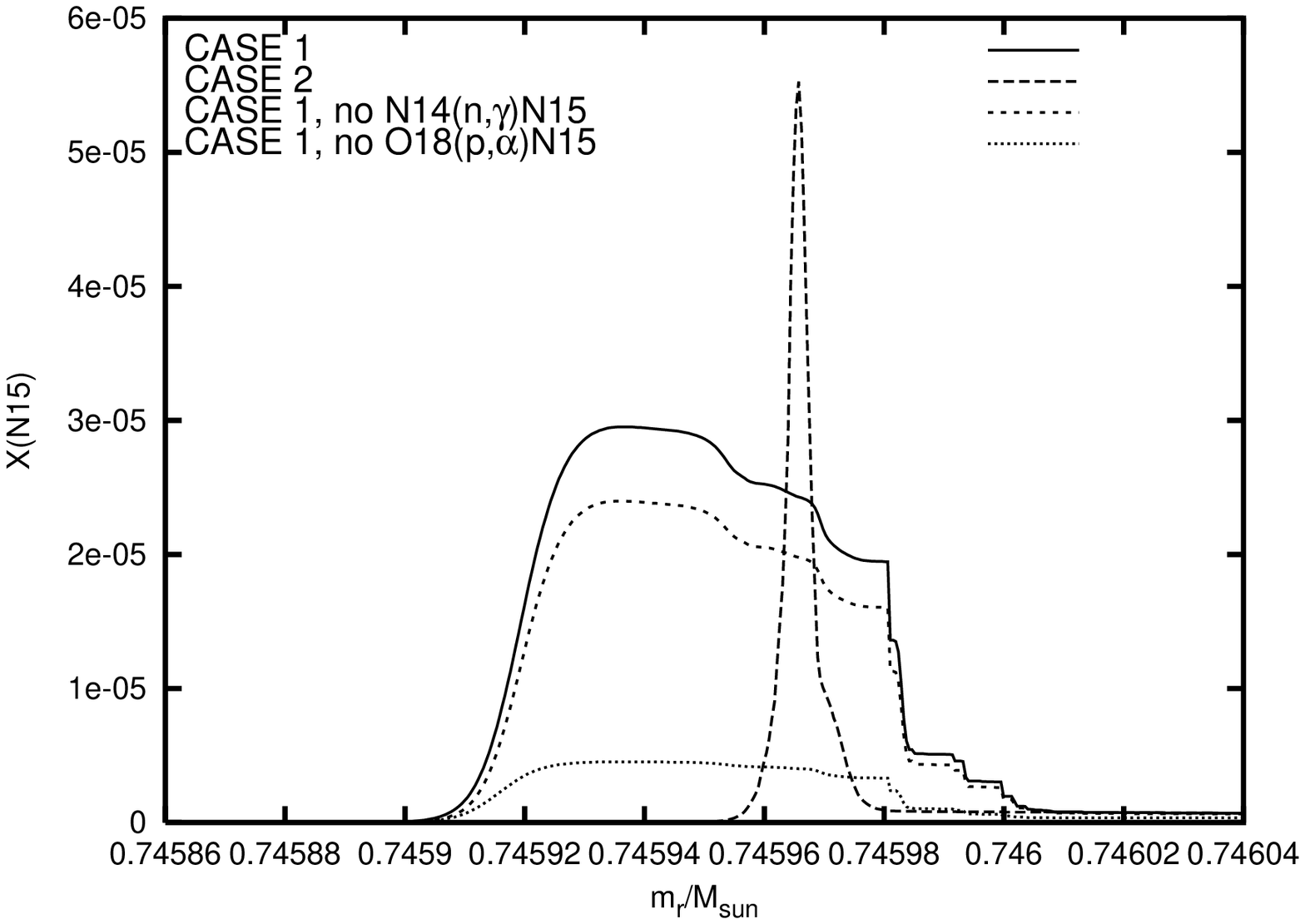}   
\caption{Maximum neutron exposure during the interpulse phase (left panel)
and final $\nfu$-abundance (right panel) for the two cases described in
the
text.}
\end{figure}
For comparison, we follow
the nucleosynthesis without mixing during the interpulse phase, i.e.
we  only process
the $\cdr$-pocket which has formed due to rotationally induced mixing
immediately after the third dredge up \textbf{[CASE 2]}.
Such a model is conceptually equivalent to models without
rotational mixing as discussed, e.g., in Lugaro \& Herwig (2001).

We find that the
maximum neutron density achieved for \textbf{CASE 1} is about one order of
magnitude smaller than in \textbf{CASE 2},
and accordingly the peak neutron exposure is
significantly smaller (Fig.\,1, left). The production of trans-iron elements
is smaller by a factor of five as most neutrons
are captured by $\nvi$. The neutron exposure integrated over
the entire s-process layer is about 1.5 times larger, since the
$\cdr$-pocket is significantly broadened by rotation and the
neutrons are partially recycled by $\nvi(\n,\mem{p})\cvi$. Some $\nfu$ is
produced, mainly  by $\cvi(\alpha,\gamma)\oac(p,\alpha)\nfu$
(Fig.\,1, right). This $\nfu$ will be transformed in $\fne$ in the
subsequent thermal pulse phase.

Both diffusive overshoot and rotation create $\cdr$ at the beginning of the
interpulse phase. However, differential rotation persistently 
mixes the $\cdr$ and $\nvi$-rich layers during the entire interpulse phase. 
Our results indicate
that, by combining these two phenomena,
it may be possible to provide a range of trans-iron element
productions as required according to
parameterized s-process models and spectroscopic stellar
observations (Busso etal., 2001).

\noindent
\textbf{Acknowledgments:} F.H. would like to thank
D.A. VandenBerg for support through his Operating Grant from the Natural
Science and Engineering Research Council of Canada.


\begin{references}
\reference  {Busso}, M., {Gallino}, R., {Lambert}, D.~L., {Travaglio}, C., \& {Smith},
             V.~V. 2001, ApJ, 557, 802
\reference Gallino, R., Arlandini, C., Busso, M., Lugaro, M., Travaglio, C., Straniero, 
           O., Chieffi, A., \& Limongi, M. 1998, ApJ, 497, 388
\reference Herwig, F. 2000, A\&A, 360, 952
\reference Langer, N., Heger, A., Wellstein, S., \& Herwig, F. 1999, A\&A, 346, L37
\reference Lugaro, M. \& Herwig, F. 2001, Nucl. Phys. A, 688, 201, astro-ph/0010012
\end{references}
\end{document}